\documentclass[journal]{IEEEtran}  

\usepackage{cite, amssymb, epsfig, epsf, fullpage}
\usepackage[cmex10]{amsmath}

\def\bx{{\mathbf x}}
\def\bs{{\mathbf s}}
\def\ba{{\mathbf a}}

\def\bh{{\mathbf h}}
\def\bdelta{{\boldsymbol \delta}}
\def\hbx{\hat{\mathbf x}}
\def\by{{\mathbf y}}
\newcommand{\rR}[0]{\mathbb{R}}
\newcommand{\cN}[0]{\mathcal{N}}
\newcommand{\cI}[0]{\mathcal{I}}
\newcommand{\cJ}[0]{\mathcal{J}}

\newcommand{\ignore}[1]{}
\newcommand{\dif}[0]{\backslash}

\newtheorem{proposition}{Proposition}

\newtheorem{lemma}{Lemma}

\begin{document}

\author{Dmitry~M.~Malioutov,~ Sujay~R.  Sanghavi,
        and~Alan~S.~Willsky,~\IEEEmembership{Fellow,~IEEE}
        \thanks{Dmitry Malioutov (e-mail: dmal@alum.mit.edu) is at
        Microsoft Research, Cambridge, UK. Sujay Sanghavi (email:
        sanghavi@mail.utexas.edu) is with the Electrical and Computer 
		Engineering department at the University of Texas, Austin.
		Alan Willsky is with the Department of Electrical
        Engineering and Computer Science, Massachusetts Institute of
        Technology, Cambridge, MA.}  \thanks{PREPRINT. The article is to appear in 
		IEEE Transactions on Special Topics in Signal Processing. This work was 
		supported
        by the Army Research Office under Grant W911NF-05-1-0207, and
        the Air Force Office of Scientific Research under Grant
        FA9550-04-1-0351.} }       

\title{ Sequential Compressed Sensing}

\maketitle

\begin{abstract}
Compressed sensing allows perfect recovery of sparse signals (or
signals sparse in some basis) using only a small number of random
measurements. Existing results in compressed sensing literature
have focused on characterizing the achievable performance by bounding 
the number of samples required for a given level of signal sparsity.
However, using these bounds to minimize the number of samples requires 
a-priori knowledge of the sparsity of the unknown signal, or the 
decay structure for near-sparse signals. Furthermore, there are some 
popular recovery methods for which no such bounds are known.

In this paper, we investigate an alternative scenario where
observations are available in sequence. For any recovery method, this
means that there is now a sequence of candidate reconstructions. We
propose a method to estimate the reconstruction error directly from the 
samples themselves, for every candidate in this sequence. This estimate is
universal in the sense that it is based only on the measurement
ensemble, and not on the recovery method or any assumed level of sparsity
of the unknown signal. With these estimates, one can now stop
observations as soon as there is reasonable certainty of either exact
or sufficiently accurate reconstruction. They also provide a way to
obtain ``run-time'' guarantees for recovery methods that otherwise
lack a-priori performance bounds.

We investigate both continuous (e.g. Gaussian) and discrete
(e.g. Bernoulli) random measurement ensembles, both for
exactly sparse and general near-sparse signals, and with both noisy 
and noiseless measurements.
\end{abstract}


\IEEEpeerreviewmaketitle

\section{Introduction}
\label{sec:intro}

In compressed sensing (CS) \cite{Candes:compressive_sampling, Donoho_CS} 
a few random linear measurements of a signal are taken, and the signal is
recovered using the additional knowledge that either the signal
or some linear transform of it is sparse. These ideas have 
generated a lot of excitement in the signal processing and machine 
learning communities, and have been applied to a range of applications 
such as magnetic resonance imaging (MRI) \cite{Lustig_MRI}, computational 
photography \cite{Baraniuk_single_pixel}, wireless networks 
\cite{wireless_CS}, and structure discovery in biological networks 
\cite{Seeger_bio}.

The applications where compressed sensing is most beneficial (e.g. MRI) 
have a high cost of acquiring each additional sample. If this cost (in terms 
of time, power, e.t.c) is high as compared to the cost of computation, then 
it is suitable to use sophisticated recovery algorithms which include the 
$\ell_1$-based \emph{basis pursuit}~\cite{Chen_Donoho:basis_pursuit}, greedy 
approaches \cite{Tropp_greedy}, and even non-convex ($\ell_p$) or iterative 
formulations \cite{Wotao_Yin_lp, Cetin:lp_ICASSP, reweighted_l1_Wakin} to 
enable recovery from fewer measurements.

While some of the recovery methods, especially those based on
$\ell_1$-regularization, have analytically provable performance
guarantees \cite{Candes_Romberg_Tao, Donoho_CS}, others, such as
non-convex $\ell_p$, reweighted $\ell_1$ \cite{reweighted_l1_Wakin},
and sparse Bayesian learning (SBL) \cite{Wipf_SBL} do not, and they
have been shown empirically to often require even fewer samples than
$\ell_1$-based methods. Furthermore, when guarantees do exist, they
have been empirically observed to sometimes be highly pessimistic 
and may require large dimensions to hold with high probability
\cite{Candes:compressive_sampling, Rud_Ver:CS}. Another drawback is 
that much of the existing analysis characterizes how many measurements 
are needed for a signal with a given sparsity level. However, as the 
sparsity level is often not known a-priori, it can be very challenging 
to use these results in practical settings.

In this paper we take an alternative approach and we develop estimates
and bounds for the reconstruction error using only the observations,
without any a-priori assumptions on signal sparsity, or on the
reconstruction method. We consider a scenario where one is able to get
observations in sequence, and perform computations in between
observations to decide whether enough samples have been obtained --
thus allowing to recover the signal either exactly or to a given
tolerance from the smallest possible number of random observations. 
This, however, requires a computationally efficient approach to detect
exactly when enough samples have been received. To get an intuition
behind our approach -- suppose that we first attempt to reconstruct
the signal while withholding some available observations, akin to
cross-validation. The observations correspond to a known linear
function of the true signal, so if the reconstructed signal is quite
different from the true signal, then the same linear function applied
to our recovered signal will result in a value that is far from the
actual observation, with high probability. Our results provide
estimates of the reconstruction error based on the statistics of the
measurement model.  They can thus be used to provide 'run-time'
guarantees even for decoders that are otherwise not amenable to
analysis.

We first consider the case when noiseless measurements are taken using the random 
Gaussian (or generic continuous) ensemble, and we show that simply checking for 
one-step agreement provides a way to check exactly when enough samples have been 
received. Suppose that after receiving $M$ samples $y_i = \ba_i' \bx,~i=1,..,M$, 
we apply a sparse reconstruction method of our choice, and obtain a solution 
$\hbx^M$ satisfying all the $M$ measurements. We can use any sparse 
decoder, including greedy matching pursuit, SBL, $\ell_p$ formulations, and even 
the brute-force decoder, but we require that the solution at each step $M$ 
satisfies $y_i = \ba_i' \hbx^M$, for $i=1,..,M$. For example, in the case of basis pursuit, we 
would solve 
\begin{equation}
\label{eqn:seq_lp}
\hbx^M = \arg \min ||\bx||_1 ~~~\mbox{s.t.}~~~ \ba_i' \bx = y_i,~~i=1,..,M.
\end{equation}
Next, we receive one more measurement, and check for one step agreement: 
i.e. if $\hbx^{M+1} = \hbx^{M}$, then the
decoder declares $\hbx^M$ to be the reconstruction and stops
requesting new measurements. In Section \ref{S:Gauss_results} we show
in Propositions \ref{prop:Gauss_new_meas} and \ref{prop:Gauss_nonsing}
that this decoder gives exact reconstruction with probability one.

For some other measurement ensembles, such as random Bernoulli and the
ensemble of random rows from a Fourier basis, the one-step agreement
stopping rule no longer has zero probability of error. We modify the
rule to wait until $T$ subsequent solutions $\hbx^M$, ..., $\hbx^{M +
T}$ all agree. In Section \ref{S:binary} we show in Proposition 
\ref{prop:bin_new_meas} that in the Bernoulli case the probability of 
making an error using this stopping rule decays exponentially with $T$, 
allowing trade-off of error probability and delay. 

In Sections \ref{S:near_sparse} and \ref{S:noisy} we show how the
error in reconstruction can be estimated from the sequence of
recovered solutions.  We first present analysis for the Gaussian
measurement ensemble in Proposition \ref{prop:near_sparse}, and then
generalize to any sensing matrices with i.i.d.  entries. This enables
the decoder to stop once the error is below a required tolerance --
even for signals that are not exactly sparse, but in which the energy
is largely concentrated in a few components, or for measurements which
are corrupted by noise.

Finally, in Section \ref{S:lin_prog} we motivate the need for efficient
solvers in the sequential setting. We consider the basis pursuit sparse 
solver and show that rather than re-solving the problem from scratch 
after an additional measurement is received, we could use an
augmented linear program that uses the solution at step $M$ to guide
its search for the new solution. We show empirically that this
approach significantly reduces computational complexity.

During the review process we learned about a very recent analysis in
\cite{Ward} for the cross-validation setting, using the 
Johnson-Lindenstrauss lemma. We describe similarities and differences 
from our work in the discussion in Section \ref{S:near_sparse}. Our
current paper extends our earlier results presented in \cite{ICASSP08:seq_cs}.


\section{Brief overview of compressed sensing}
\label{S:CS_overview}

As there is no dearth of excellent tutorials on compressed sensing 
\cite{Baraniuk_CS, Candes:compressive_sampling, Donoho_CS}, in 
this section we give only a brief outline mainly to set the stage for 
the rest of the paper. At the heart of compressed sensing lies the 
sparse recovery problem\footnote{ The ground-breaking results \cite{Donoho_sp_rep} 
predating compressed sensing were in context of sparse signal representation where 
one seeks to represent a vector $\by$ in an overcomplete dictionary 
$A \in \rR^{M \times N}$, $M << N$, with coefficients $\bx$, i.e., 
$\by = A \bx$.}, which tries to reconstruct an unknown sparse 
signal $\bx$ from a limited number of measurements $\by = A \bx$, where 
$A \in \rR^{M \times N}$, $M << N$.  Much of excitement in the field 
stems from the fact that the hard combinatorial 
problem of searching for sparse solutions in the affine space $\{ \bx: \by = 
A \bx\}$ under certain suitable conditions can be solved exactly via 
various tractable methods. The most widely known methods include greedy matching 
pursuit and its variants \cite{Tropp_greedy}, and approaches based on convex 
optimization, using $\ell_1$ norms as a proxy for sparsity 
\cite{Chen_Donoho:basis_pursuit}:
\begin{equation}
  \label{eqn:simple_sparse_rep}
  \min \Vert \bx \Vert_1 ~~~\mbox{ subject to }~~ \by = A \bx.
\end{equation}

An early sufficient condition for sparse recovery 
\cite{Donoho_sp_rep} states that the formulation in (\ref{eqn:simple_sparse_rep}) 
recovers the unique sparse solution if $A$ is well-posed and $\bx$ is 
sparse enough, i.e. if $\Vert \bx \Vert_0 < \frac{1 + 1/M(A)}{2}$, where 
$M(A) = \max_{i \ne j} | \ba_i' \ba_j |$, and $A$ has columns $\ba_i$
normalized to $1$. However, this simple condition is very pessimistic. 
Much tighter conditions are obtained by considering larger subsets
of columns of $A$, e.g. the restricted isometry property (RIP) depends on 
the maximum and minimum singular values over all $M \times K$ submatrices
of $A$ \cite{Candes_Romberg_Tao}. Namely, a matrix $A$ satisfies the $K$-RIP
with constant $\delta_K	$ if $(1 - \delta_K) \Vert \bx \Vert_2^2 \le
\Vert A \bx \Vert_2^2 \le (1 + \delta_K) \Vert \bx \Vert_2^2$ for every 
$\bx$ which has at most $K$ non-zero entries. While enabling much 
tighter sufficient conditions for recovery of sparse signals 
\cite{Candes_Romberg_Tao}, the RIP is very  costly (exponential in $K$) 
to check for a given matrix. 

Results in compressed sensing take advantage of RIP by bringing in the theory of 
random matrices into the picture. In compressed sensing we receive
random measurements $\by = \Psi \bs$ where the unknown signal of interest $\bs$ is 
itself sparse in some basis, i.e. $\bs = \Phi \bx$, with $\bx$ sparse. Hence 
the problem reduces to finding sparse solutions satisfying $\by = \Psi \Phi \bx = A \bx$, 
where $A = \Psi \Phi$ is a random matrix. 

A collection of results have been established that RIP holds for random 
matrices of certain size from given ensembles: Gaussian, Bernoulli, random 
Fourier rows \cite{Candes_Romberg_Tao, Donoho_CS, Rud_Ver:CS}. The general 
conclusion of these results is that the convex $\ell_1$ formulation can 
recover (with high probability) a signal $\bx \in \rR^N$ with $K$ non-zeros 
from only $C K \log(N)$ measurements, where $C$ is a constant depending on 
the random measurement ensemble. This is indeed remarkable -- as it only 
requires a logarithmic dependence of the number of measurements on $N$.

\begin{figure}
  \begin{center}
    \centerline{\epsfig{figure=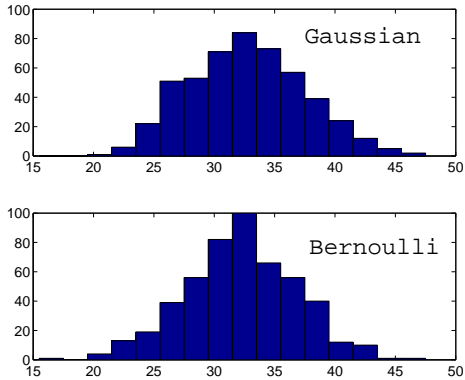,width=2.85in}}
    \caption{\label{fig:stop_times} Histogram of the stopping times distribution 
	for Gaussian and Bernoulli measurement ensembles: $N = 100$, and $K=10$, and 
	$\ell_1$ decoding.}
  \end{center}
\end{figure}

However, when each additional measurement is very costly there 
are several problems with these bounds -- firstly, since they are 
high-probability results independent of $\by$, they tend to be conservative, 
and also the constants $C$ are typically generous upper-bounds. Secondly, the 
number of measurements depends on the number of non-zero components of $\bx$ 
which may not be known a-priori. Finally, there are successful approaches which 
we mentioned in Section \ref{sec:intro} for which no such results are available.

In Figure \ref{fig:stop_times} we illustrate the drawbacks of using upper bounds 
on the number of measurements. We find the minimum number $M$ of random samples which 
were needed to recover a sparse signal $\bx$ with $N = 100$, and $K = 10$ 
from random Gaussian and Bernoulli measurements using the $\ell_1$-formulation in 
(\ref{eqn:simple_sparse_rep}), over $500$ random trials.  We plot a histogram of 
these numbers, and we see that they exhibit high variance, and so relying on 
conditions that guarantee recovery with high probability often means taking many 
unnecessary samples. This motivates the need for sequential compressed sensing 
scenario that can adaptively minimize the number of samples for each observed $\by$, 
which we describe next. 


\section{Stopping rule in the noiseless continuous case}
\label{S:Gauss_results}

We now analyze the sequential CS approach for the case when the measurements vectors
$\ba_i$ come from a continuous ensemble (e.g., the i.i.d. Gaussian ensemble),
having the property that with probability $1$ a new vector $\ba_{M+1}$ will not be in any 
lower-dimensional subspace determined by previous vectors
$\{ \ba_i \}_{i = 1}^M$. Suppose that the underlying sparse 
signal $\bx^* \in \rR^N$ has $K$ non-zero components (we denote the number 
of non-zero entries in $\bx$ by $\Vert \bx \Vert_0$). We sequentially receive 
random measurements $y_i = \ba_i' \bx^*$, where for concreteness
$\ba_i \sim \cN(0, I)$ is a $N$-vector of i.i.d. Gaussian samples, but 
the analysis also holds if entries of $\ba_i$ are i.i.d. samples of an arbitrary 
continuous random variable.  At step $M$ we use a sparse solver of our choice 
to obtain a {\em feasible}\footnote{This requirement is essential for the noiseless
case (it is relaxed in later sections). For greedy methods such as matching pursuit 
this means that we allow enough iterations until all the measurements received so far
are satisfied perfectly. Noiseless basis pursuit formulations satisfy it by construction.} 
solution $\hbx^M$ using all the received data. Results in compressed sensing 
\cite{Candes:compressive_sampling, Rud_Ver:CS} indicate 
that if we use basis pursuit or matching pursuit methods, then after receiving 
around $M \propto K \log(N)$ measurements we can recover the signal $\bx^*$ with 
high probability. This requires the 
knowledge of $K$, which may not be available, and only rough bounds on the 
scaling constants are known. Our approach is different -- we compare the 
solutions at step $M$ and $M+1$, and if they agree, we declare correct 
recovery.

\begin{figure}[t]
\centering
\input{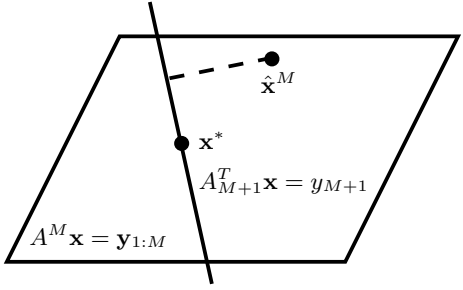}
\caption{\label{fig:random_hyperplane} A new constraint is added:
$\ba_{M+1}' \bx = y_{M+1}$. Probability that this hyperplane passing
through $\bx^*$ also passes through $\hbx^{M}$ is zero. }
\vspace{-.2cm}
\end{figure}

\begin{proposition}
\label{prop:Gauss_new_meas}
  If in the Gaussian (generic continuous) measurement ensemble it 
holds that $\hbx^{M+1} = \hbx^{M}$, then $\hbx^M = \bx^*$, with 
probability $1$.
\end{proposition}

{\em Proof.} Let $\by_{1:M} \triangleq [y_1, ..., y_M]'$, and $A^M
\triangleq [\ba_1, ..., \ba_M]'$. Suppose that $\hbx^M \ne \bx^*$. We
have that $\by_{1:M} = A^M \hbx^M$ and $\by_{1:M} = A^M \bx^*$: both
$\bx^*$ and $\hbx^M$ belong to the $(N-M)$-dimensional affine space
$\{\bx ~|~ \by_{1:M} = A^M \bx \}$.  The next measurement passes a
random hyperplane $y_{M+1} = \ba_{M+1}' \bx^*$ through $\bx^*$ and
reduces the dimension of the affine subspace of feasible solutions by
$1$. In order for $\hbx^M$ to remain feasible at step $M+1$, it must
hold that $\by_{M+1} = \ba_{M+1}' \hbx^M$. Since we also have
$\by_{M+1} = \ba_{M+1}' \bx^*$, then $\hbx^M$ remains feasible only if
$(\hbx^M - \bx^*)' \ba_{M+1} = 0$, i.e. if $\ba_{M+1}$ falls in the
$N-1$ dimensional subspace of $\rR^N$ corresponding to $Null((\hbx^M -
\bx^*)')$. As $\ba_{M+1}$ is random and independent of $\hbx^M$ and of
the previous samples $\ba_1$, ..., $\ba_M$, the probability that this
happens is $0$ (event with measure zero). See Figure
\ref{fig:random_hyperplane} for illustration. $\square$\\

Note that the proof implies that we can simplify the decoder to checking 
whether $\ba_{M+1}' \hbx^M = y_{M+1}$, avoiding the need to solve for 
$\hbx^{M+1}$ at the last step\footnote{We thank the anonymous reviewer for
this simplification.}. Moreover, if using any sparse solver in the 
continuous ensemble case the solution $\hbx^M$ has fewer than $M$ non-zero 
entries, then $\hbx^M = \bx^*$ with probability 1.

\begin{proposition} 
\label{prop:Gauss_nonsing}
For a Gaussian (continuous) measurement ensemble, if $\Vert \hbx^M \Vert_0 < M$, 
then $\hbx^M = \bx^*$ with probability $1$.\footnote{Note that a random 
measurement model is essential:
for a fixed matrix $A$ if $2 K > M$ then there exist $\bx_1$ and
$\bx_2$ such that $A \bx_1 = A \bx_2$ and $\Vert \bx_i \Vert_0 \le
K$. However, for a fixed $\bx^*$ with $\Vert \bx^* \Vert_0 < M$ the
probability that it will have ambiguous sparse solutions for a random
choice of $A$ is zero.}
\end{proposition}

{\em Proof.} Denote the support of our unknown sparse vector $\bx^*$ by $\cI$, 
i.e. $\cI = \{i ~|~ x^*_i \ne 0\}$. We next generate a random measurement 
matrix $A^M$. Let $A = A^M$ to simplify notation. We receive the 
corresponding measurements $\by = A \bx^*$. Now $A$ is $M \times N$, with 
$M < N$. The key fact about random matrices 
with i.i.d. entries from a continuous distribution is that any $M \times M$ 
submatrix of $A$ is non-singular with probability $1$\footnote{This is easy 
to see: fix $T \subset \{1, ..., N\}$ with $|T| = M$. Then probability that 
$A_{T_M} \in span(A_{T_1}, ...,A_{T_{M-1}})$ is zero, as $A_{T_M}$ is 
a random vector in $\rR^M$ and the remaining columns span a 
lower-dimensional subspace.}. We now argue that with probability 1 after 
receiving $\by$ there will not exist another sparse feasible 
solution $\hbx \ne \bx^*$, i.e. $\hbx$ with fewer than $M$ non-zero 
entries satisfying $\by = A \hbx$ . We consider all possible sparse supports 
$\cJ \subset \{1, .., N\}$, with $|\cJ| < M$, and show that a
feasible solution $\hbx \ne \bx^*$ can have this support only with probability 0. 
There are two cases: $\cI \subset \cJ$ and $\cI \ne \cI \cap \cJ$. 
\begin{figure}
  \begin{center}
    \centerline{\epsfig{figure=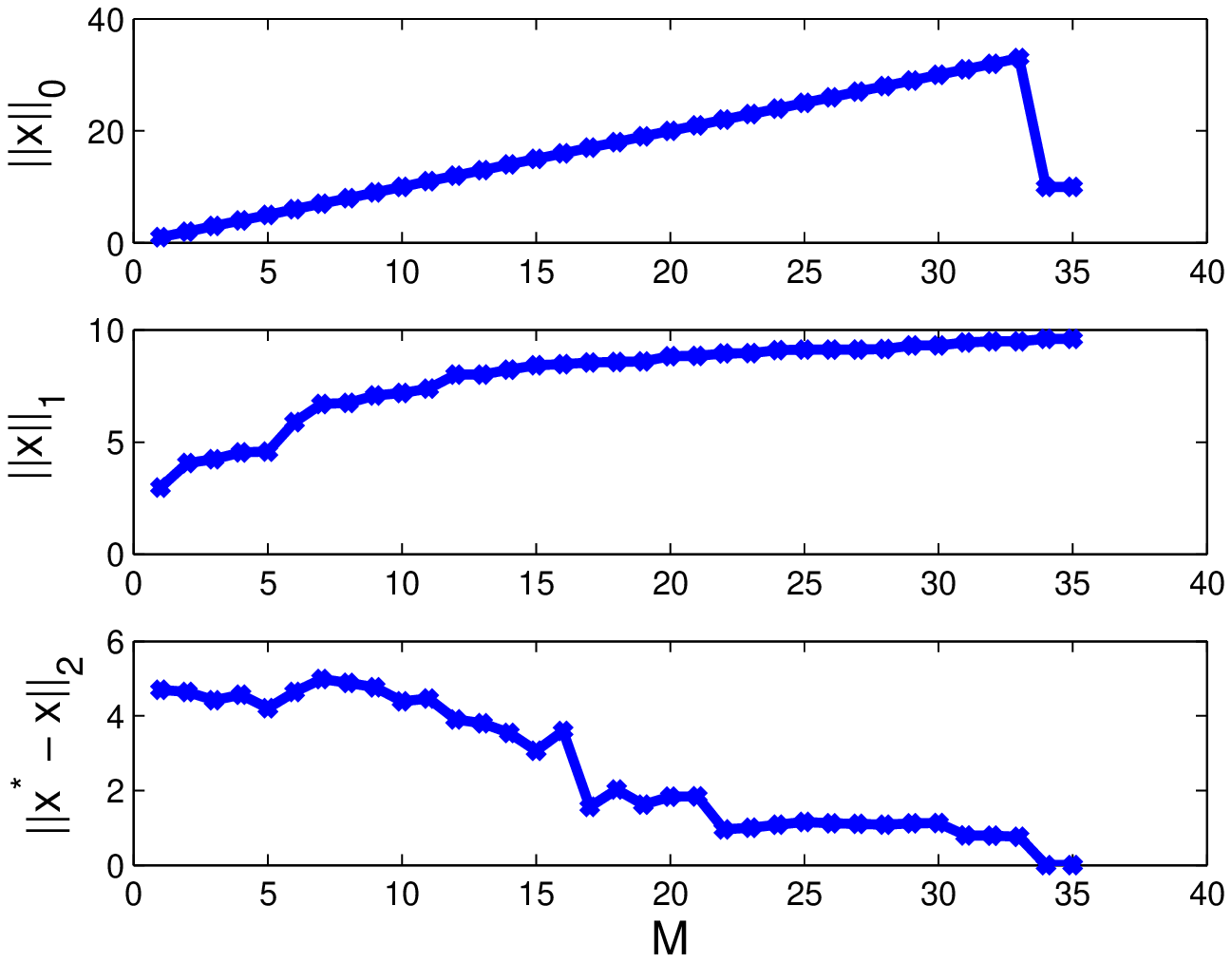,width=3in}}
    \caption{\label{fig:gauss_plot} Gaussian ensemble example: $N
    = 100$, and $K = 10$. (Top): $~\Vert \hbx^{M} \Vert_0$. (Middle):
    $~\Vert \hbx^{M} \Vert_1$. (Bottom): $~\Vert \bx^* - \hbx^{M}
    \Vert_2$.}
  \end{center}
\end{figure}

First suppose $\cI \subset \cJ$, $|\cJ| < M$, and suppose there exists some feasible 
$\hbx$ supported on $\cJ$. Then $\hbx - \bx^* \in Null(A)$, and support of $\hbx - \bx^*$ 
is a subset of $\cJ$, hence it is smaller than $M$. But that means that there is a subset 
of fewer than $M$ columns of $A$ that are linearly dependent, which can only happen with 
probability zero.

Now consider the case $\cI \ne \cI \cap \cJ$. For a fixed $\cI$ we consider 
all such possible sets $\cJ$, with $|\cJ| < M$. First fix one such set
$\cJ$. We use the notation $\cI \dif \cJ = \{ ~i \in I ~|~ i \notin
J\}$. Note that we have $\by = A \bx^* = A_{\cJ} \bx^*_{\cJ} + A_{\cI \dif \cJ}
\bx^*_{\cI \dif \cJ}$. Let $\tilde{\by} = A_{\cI \dif \cJ} \bx^*_{\cI \dif \cJ}$.  
Now since we require $\hbx$ to be feasible, we also need 
$\by = A \hbx = A_{\cJ} \hbx_{\cJ}$ which would imply that 
$\tilde{\by} = A_{\cJ} (\hbx_{\cJ} - \bx^*_{\cJ} )$. This means that 
the vector $\tilde{\by}$ would also have to be in the span of
$A_{\cJ}$. However, $\tilde{\by}$ is a random vector in $\rR^M$ (determined by
$\bx^*$ and $A_{\cI \dif \cJ}$), and span of $A_{\cJ}$ is an independent
random subspace of dimension strictly less than $M$. Hence, the event that 
$\tilde{\by}$ also falls in the span of $A_{\cJ}$ has measure zero. This means
that for a fixed $\cJ$ a distinct sparse solution can only exist with probability 
$0$. Now the number of possible subsets $\cJ$ is finite (albeit large), so even 
when we take all such supports $\cJ$, a distinct sparse solution supported on $\cJ$ 
can only exist with probability $0$. Hence, with probability $1$ there is only one 
solution with $\Vert \bx \Vert_0 < M$, namely $\bx^*$. $\square$\\

This proposition allows to stop making measurements when a feasible solution has 
less than $M$ nonzero entries -- avoiding the need to make the last $(M+1)$-st 
measurement.

Consider an example in Figure \ref{fig:gauss_plot} with $N = 100$, 
and $K = 10$. We keep receiving additional measurements and solving 
(\ref{eqn:seq_lp}) until we reach one-step agreement, $\hbx^{M} = \hbx^{M+1}$. 
The top plot shows that $\Vert \hbx^{M} \Vert_0$ increases linearly with $M$ 
until one step agreement occurs at $M = 35$, at which point it drops to $K=10$ 
and a and we recover the correct sparse solution, $\hbx^M = \bx^*$. The middle plot 
shows the monotonic increase in $\Vert \hbx^M \Vert_1$ (as the feasible set is 
shrinking with $M$). The bottom plot shows the error-norm of the solution, 
$\Vert \hbx^M - \bx^* \Vert_2$. On average it tends to go down with 
more observations, but non-monotonically. After $M=35$ the error becomes 
zero. We see that in the ideal conditions of no measurement noise, 
sparse unknown signals and Gaussian measurement ensembles, the number
of measurements can be indeed minimized by a simple stopping rule. 

\section{Stopping rule in the Bernoulli case}
\label{S:binary}

In this section we study a simple but popular measurement ensemble 
that is not one of the generic continuous ensembles described in the 
previous section.
Suppose that the measurement vectors $\ba_i$ have equiprobable
i.i.d.$~$Bernoulli entries $\pm 1$. A difference emerges from the
Gaussian case: the probability that all $M \times M$ submatrices of
$A^M$ are non-singular is no longer $0$. This makes it possible (with
non-zero probability) for $\hbx^{M+1}$ to agree with $\hbx^{M}$ even 
though $\hbx^M \ne \bx^*$, and for erroneous solutions $\hbx^M$ to have
cardinality less than $M$. We modify the stopping rule to require
agreement for several steps - success is declared only when last $T$
solutions all agree. We show in proposition \ref{prop:bin_new_meas}
that the probability of error decays exponentially with $T$. We use
the following Lemma from \cite{Tao_sing_bin_mat}:
\begin{lemma}[Tao and Vu]
\label{lem:Tao}
Let $\ba$ be an i.i.d. equiprobable Bernoulli vector with $\ba \in
\{-1, 1\}^N$. Let $W$ be a deterministic $d$-dimensional subspace of
$\rR^N$, $0\le d < N$.  Then $P( \ba \in W) \le 2^{d - N}$.
\end{lemma}

We are now ready to establish the following claim:

\begin{proposition}
  \label{prop:bin_new_meas}
  Consider the Bernoulli measurement case. If $\hbx^{M} = \hbx^{M+1} =
  ... = \hbx^{M+T}$, then $\hbx^M = \bx^*$ with probability greater
  than or equal to $1 - 2^{-T}$.
\end{proposition}

{\em Proof.} Suppose that $\hbx^M \ne \bx^*$. Denote the support of
$\bx^*$ by $\cI$ and the support of $\hbx^M$ by $\cJ$. At step $M$ we
have $A^M \bx^* = A^M \hbx^M$. Let $W = \{ \ba ~|~ (\hbx^M - \bx^*)'
\ba = 0 \}$, i.e. the nullspace of $(\hbx^M - \bx^*)'$. Then $W$ is an
$(N-1)$-dimensional subspace of $\rR^N$.

Given a new random Bernoulli sample $\ba_{M+1}$, the vector $\hbx^{M}$
can remain feasible at step $M+1$ only if $(\hbx^M - \bx^*)'
~\ba_{M+1} = 0$, i.e. if $\ba_{M+1}$ falls into $W$. By Lemma
\ref{lem:Tao}, the probability that $\ba_{M+1} \in W$ is a most
$1/2$. The same argument applies to all subsequent samples of
$\ba_{M+i}$ for $i = 1, .., T$, so the probability of having $T$-step
agreement with an incorrect solution is bounded above by
$2^{-T}$. $\square$\\

Note that as in the discussion for the continuous case, we can simply
check that $\ba_{M+i}' \hbx^M = y_{M+i}$ for $i = 1,..., T$, avoiding the 
need to solve for $\hbx^{M+T}$.

We now pursue an alternative heuristic analysis, more akin to
Proposition \ref{prop:Gauss_nonsing}. For the Bernoulli case, $\Vert
\hbx^M \Vert_0 < M$ does not imply $\hbx^M = \bx^*$. However, we
believe that once we obtain enough samples so that $N^2 2^{1- M}$ $\ll 1$
then $\Vert \hbx^M \Vert_0 < M$ will imply that $\hbx^M = \bx^*$ with 
high probability. Since the elements of $\ba_i$ belong to finite set 
$\{-1, 1\}$, an $M \times M$ submatrix of $A^M$ can be singular with
non-zero probability. Surprisingly, characterizing this probability is
a very hard question.  It is conjectured \cite{Tao_sing_bin_mat} that
the dominant source of singularity is the event that two columns or
two rows are equal or opposite in sign. This leads to the following
estimate (here $X_M$ is $M \times M$):\footnote{Probability that two
columns are equal or opposite in sign is $2^{1-M}$, and there are
$O(M^2)$ pairs of columns.}
\begin{equation}
  P( \det X_M = 0) = (1 + o(1)) M^2 2^{1- M}.
\end{equation}
However the very recent best provable bound on this probability is
still rather far: $P(\det X_M = 0) = ((\frac{3}{4} +
o(1))^M)$~\cite{Tao_sing_bin_mat}. If we assume that the simple
estimate based on pairs of columns is accurate, similar analysis shows
that the probability that a random $\pm 1$ $M \times N$ matrix with $M
\ll N$ having all $M \times M$ submatrices non-singular is $(1 + o(1))
N^2 2^{1- M}$.

\section{Near-sparse signals}
\label{S:near_sparse}

In practical settings, e.g. when taking Fourier and wavelet transforms
of smooth signals, we may only have approximate sparseness: a few
values are large, and most are very small. In this section we extend
our approach to this case; again, and in contrast to existing work,
we do not need to assume a specific near-sparse structure, like
power-law decay, but instead provide bounds that hold for any signal.

\begin{figure}[t]
\centering
\input{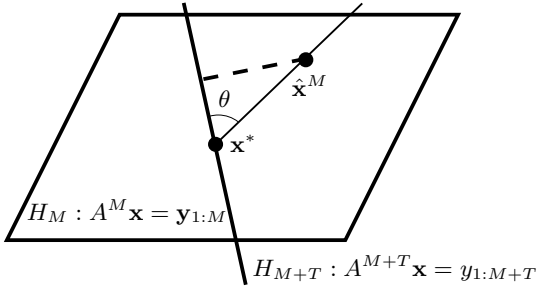}
\caption{\label{fig:random_hyperplane_near_sparse} Geometry of the analysis for 
near-sparse signals. The unknown reconstruction error is related to 
$d( \hbx^M, H_{M+T})$ and the angle $\theta$ between the line from
$\bx^*$ to $\hbx^M$ and the affine space $H_{M+T}$ defined by the new 
measurements.}
\vspace{-.2cm}
\end{figure}

The exact one-step agreement stopping rule from Section
\ref{S:Gauss_results} is vacuous for near-sparse signals, as $\Vert
\bx^* \Vert_0 = N$, and all samples are needed for perfect
recovery. We start by considering Gaussian measurements, and show that
we can gather information about the current reconstruction error by obtaining 
a small number of additional measurements, and computing the distance between
the current reconstruction and the affine space determined by these
new measurements. The reconstruction error is then equal to an 
unknown constant times this distance:
\begin{equation}
  \label{eqn:near_sparse_error}
   \Vert \bx^* - \hbx^M \Vert_2 = C_T ~d( \hbx^M, H_{M+T}),
\end{equation}
where $H_{M+T} \triangleq \{ x ~|~ y_i = \ba_{i}' x, ~1 \le i \le
M+T\}$ is the affine space determined by all $M+T$ measurements, 
$C_T$ is a random variable that we will bound, and $d( \hbx^M, H_{M+T})$ 
denotes the distance from $\hbx^M$ to $H_{M+T}$. We characterize $E[C_T]$ and
$Var[C_T]$ -- this gives us a confidence interval on the
reconstruction error using the observed distance $d( \hbx^M,
H_{M+T})$. We can now stop taking new measurements once the error
falls below a desired tolerance. Note that our analysis does not
assume a model of decay, and bounds the reconstruction error by
obtaining a small number of additional measurements, and computing the 
prediction error. In contrast, some related results in CS
literature assume a power-law decay of entries of $\bx^*$ (upon
sorting) and show that with roughly $O(K \log N)$ samples, $\hbx^M$ in
(\ref{eqn:seq_lp}) will have similar error to that of keeping the $K$
largest entries in $\bx^*$ \cite{Candes:compressive_sampling}.

We now outline the analysis leading to a bound based on (\ref{eqn:near_sparse_error}).  
Consider Figure \ref{fig:random_hyperplane_near_sparse}. 
Let $H_M = \{ \bx : A^M \bx  = \by_{1:M} \}$ be 
the subspace of feasible solutions after $M$ measurements. Both $\bx^*$ and $\hbx^M$ lie
in $H_M$. The affine space $H_{M+T}$ is contained in $H_M$. Let $L = N - M$, and 
$\theta_T$ be the angle between the vector $\hbx^M - \bx^*$ and the affine space $H_{M+T}$.
Both are contained in the $L$-dimensional space $H_M$. Centering around $\bx^*$, we see that
$\theta_T$ is the angle between a fixed vector in $R^L$ and a random $L - T$ 
dimensional subspace of $R^L$, and the constant $C_T$ in (\ref{eqn:near_sparse_error}) is
equal to $\frac{1}{\sin(\theta_T)}$:
\begin{equation}
  \label{eqn:near_sparse_error_v2}
  \Vert\bx^*- \hbx^M \Vert_2 = \frac{d(\hbx^M, H_{M+T})}{\sin(\theta_T)},
\end{equation}
We next analyze the distribution of $\theta_T$ and hence of $C_T$. In distribution, 
$\theta_T$ is equivalent to the angle between a fixed $L-T$ dimensional subspace, 
say the one spanned by the last $L - T$ coordinates, and an i.i.d. Gaussian vector 
(whose direction falls uniformly on a unit sphere in $\rR^L$). This holds because 
the distribution of an i.i.d. Gaussian sample does not get changed after applying an 
arbitrary orthogonal transformation. Let $H$ be the span of the last $L-T$ coordinate
vectors, and $\bh$ be i.i.d. Gaussian. Then:

\begin{equation}
  C_T = \frac{1}{\sin(\theta)} = \sqrt{\sum_{i=1}^L h_i^2} ~\mbox{\Large /}~
  \sqrt{\sum_{i=1}^T h_i^2}.
\end{equation}

Using the properties of $\chi_L$, $\chi^2_L$, and inverse-$\chi^2_L$
distributions \cite{Kotz:Stat_dist} and Jensen's inequality, we have
an estimate of the mean $E[C_T] \approx \sqrt{\frac{L}{T}}$ and an 
upper bound on both the mean and the variance:
\begin{eqnarray}
\label{eqn:mean_var_bnds}
  E\left[C_T\right] \le \sqrt{\frac{L-2}{T-2}},\\
  Var\left[C_T\right] \le \frac{L-2}{T-2} - \frac{L}{T}.
\end{eqnarray} 

We describe the analysis in Appendix \ref{App:distrib_angles}. Using these 
bounds in conjunction with the Chebyshev inequality\footnote{To improve upon 
Chebyshev bounds we could directly characterize the cumulative density function 
of $C_T$ -- either analytically, or by simple Monte Carlo estimates.}, 
$p( |a - E[a]| \ge k \sigma_a) \le \frac{1}{k^2}$, we have the following result:\\

\begin{figure}[t]
\begin{center}
\epsfig{figure=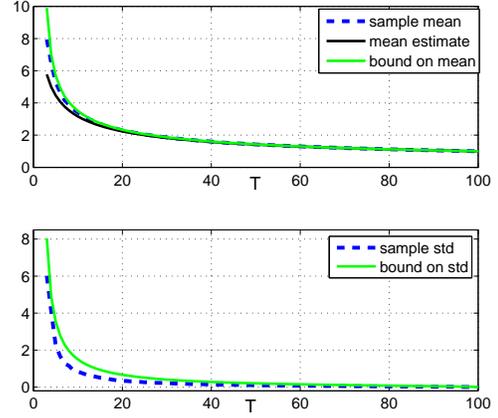,width=3in}
\caption{\label{fig:est_theta} (Top) sample mean, estimate of the mean, 
and a bound on the mean of $C_T$. (Bottom) sample standard deviation, and a 
bound on the standard deviation of $C_T$. Sample mean is based on 1000 
samples. $L = 100$.}
\end{center}
\end{figure}

\begin{proposition}
\label{prop:near_sparse}
In the Gaussian measurement ensemble we have: $\Vert \bx^* - \hbx^M \Vert_2 
\le \bar{C}^k_T ~d(\hbx^M, H_{M+T})$ 
with probability at least $1 - \frac{1}{k^2}$, where 
$\bar{C}^k_T = \sqrt{\frac{L-2}{T-2}} + k \sqrt{ \frac{L-2}{T-2} - \frac{L}{T} }$, 
for any $k > 0$.\\ 
\end{proposition} 

\begin{figure}[t]
\begin{center}
\begin{tabular}{ c }
\parbox[l]{2.75in}{
    \centerline{\epsfig{figure=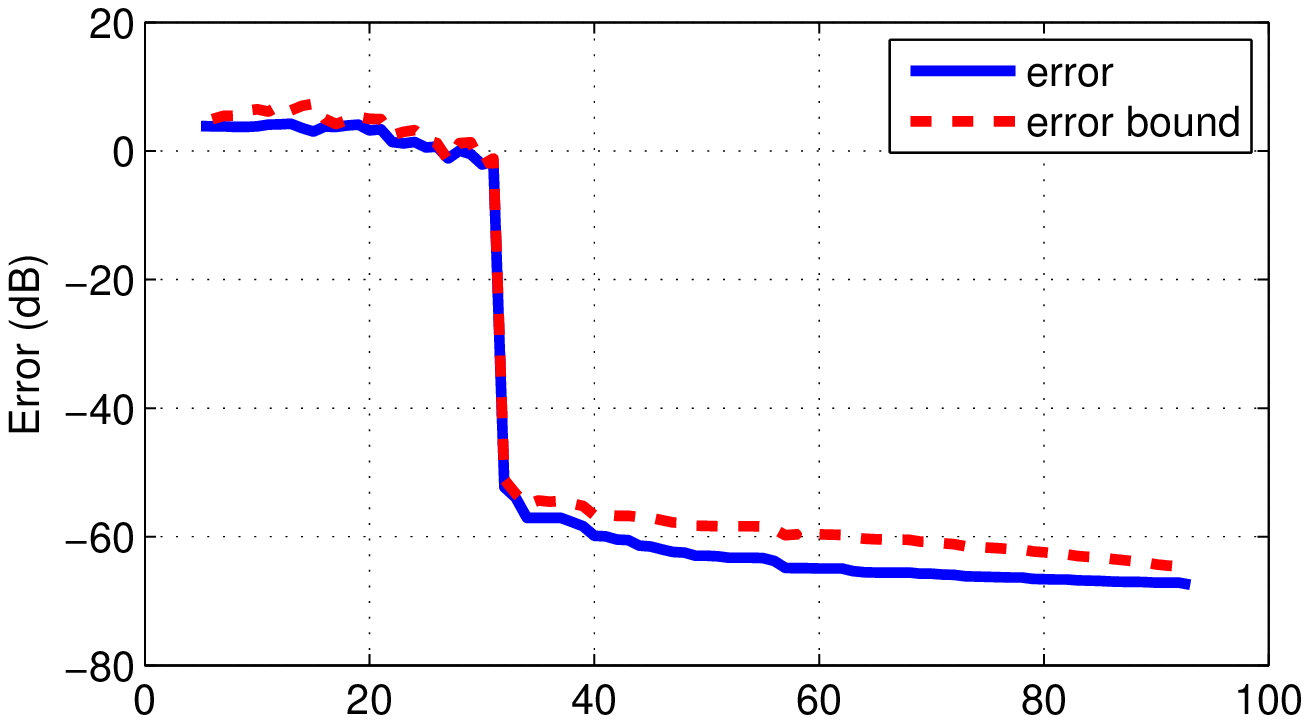,width=2.75in}}} \\
  \parbox[l]{2.75in}{
    \centerline{\epsfig{figure=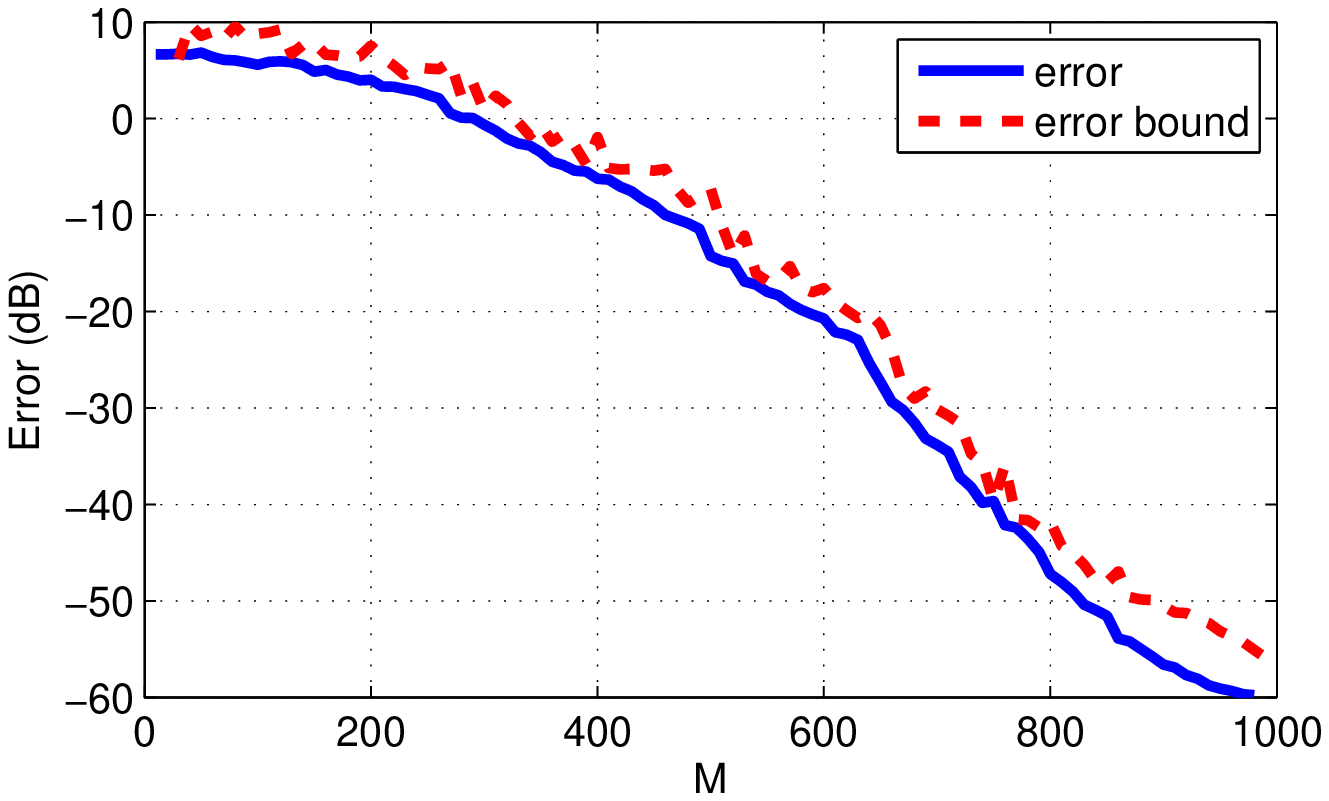,width=2.75in}}}
\end{tabular}
\caption{\label{fig:near_sparse_bnds} (Top) Error confidence bounds and actual errors
for a sparse signal, $N = 100$, $T = 5$, $K = 10$. (Bottom): Error confidence bound 
and actual errors for a signal with power-law decay, $N=1000$, $T=10$.}
\end{center}
\end{figure}

In Figure \ref{fig:est_theta} (top) we plot the mean estimate, and our bound in 
(\ref{eqn:mean_var_bnds}) for $C_T$ and (bottom) the standard deviation bound for 
$L = 100$ and a range of $T$. We compare them to sample mean and standard deviation 
of $C_T$ based on $5000$ samples. The figure shows that both bounds provide very good 
approximation for most of the range of $T > 2$, and also that the standard deviation 
quickly falls off with $T$, giving tight confidence intervals. In Figure 
\ref{fig:near_sparse_bnds} we perform numerical experiments with two example signals, 
a sparse signal, $N= 100$, $K = 10$, $T = 5$ (top) and a near-sparse signal with power-law 
decay, $N = 1000$, $T = 10$ (bottom). We use basis pursuit to recover the signals as we 
obtain progressively more measurements, and we compare our error bounds
(via Chebyshev inequality) to the actual errors. We see that the bounds reliably 
indicate the reconstruction error -- after a small delay of $T$ additional measurements. 
We have used basis pursuit in the experiments, but we could substitute any sparse solver instead, 
for example we could have also computed error estimates for matching pursuit. 

\subsection{Analysis for More General Ensembles}
\label{S:near_sparse_simplified}

To get the bound in (\ref{eqn:near_sparse_error}) we characterized the
distribution of $\frac{1}{\sin(\theta_T)}$ and used the properties of
the Gaussian measurement ensemble. Analysis of $\theta_T$ for general
ensembles is challenging. We now consider a simpler analysis which
provides useful estimates when $T << L$, i.e. the case of main
interest for compressed sensing, and when the measurement coefficients
$a_{ij}$ are from an i.i.d. zero-mean ensemble. The previous bound for
the Gaussian case depended on both $M$, the number of samples used for
the current reconstruction, and $T$, the number of extra
samples. Now, in the following we give estimates and bounds that 
depend only on $T$, and in that sense could be weaker for the Gaussian 
case when $M$ is large; they are however more generally applicable --
in particular we no longer require $\hbx^M$ to satisfy the measurements
exactly. 

Suppose we have a current reconstruction $\hbx$, and suppose $\bx^*$
is the (unknown) true signal. We now take $T$ new samples $y_i = \ba_i'
\bx^*$, for $1\leq i \leq T$. For each of these samples we compute
$\hat{y}_i = \ba_i' \hbx^M$ to be the {\em same} vector $\ba_i$ 
applied to the current reconstruction. Denote the current
error vector by $\bdelta = \hbx^M - \bx^*$, and compute $z_i =
\hat{y}_i - y_i$, the deviations from the actual measurements.  Then
\begin{equation}
\label{eqn:JL_analysis}
 z_i = \ba_i' \bdelta, ~~~1 \le i \le T
\end{equation}
The new measurements $\ba_{i}$ are independent of $\hbx$ and of
$\bx^*$, hence of $\bdelta$. The $z_i$'s are i.i.d. from some
(unknown) distribution, which has zero mean and variance $\Vert
\bdelta \Vert_2^2 ~Var( a_{ij})$. We can estimate $\Vert \bdelta
\Vert_2^2$ by estimating the variance of the $z_i$'s from the $T$
samples. The quality of the estimate will depend on the exact 
distribution of $\ba_{ij}$.

Consider the case where $\ba_i$ are i.i.d. Gaussian. Then $z_i$ 
is Gaussian as well. For simplicity suppose that $Var(a_{ij}) = 1$, then 
the distribution of $z_i$ is i.i.d. Gaussian with zero-mean and variance
$\Vert \bdelta \Vert_2^2$. Let $Z_T = \sum_{i = M+1}^{M+T}
z_i^2$. Then $\tilde{Z}_T \triangleq \frac{Z_T}{\Vert \delta \Vert^2}
\sim \chi^2_T$, i.e. $\chi^2$ random variable with $T$ degrees of
freedom. Now to obtain a confidence interval for $\Vert \bdelta \Vert_2^2$ we 
use the cumulative $\chi^2_T$ distribution. We pick a
confidence level $1 - \alpha$ (for some small $\alpha > 0$), and we use
the $\chi^2_T$ cumulative distribution to find the largest $z^*$ such
that $p(\tilde{Z}_T \le z^*) \le \alpha$.\footnote{ We have that
$\sigma_z^2 = \frac{Z_T}{z^*}$ gives the smallest value of $\sigma_z^2$
such that probability of observing $Z_T$ is at least $\alpha$. That is
to say, the bound $\Vert \bdelta \Vert^2 < \frac{Z_T}{z^*}$ will hold
for at least $1- \alpha$ fraction of realizations of $Z_T$.}

During the review process a related analysis in \cite{Ward} was brought to our 
attention: the paper considers compressed sensing in a cross-validation 
scenario, and it proposes to estimate the errors in the reconstruction from 
a few additional (cross-validation) measurements. The paper cleverly uses
the Johnson-Lindenstrauss (JL) lemma to find out how many random measurements are
needed for predicting the error to a desired accuracy. For Gaussian measurements
ensembles our $\chi^2$-based analysis can be seen as a special case (where all 
the constants are computed explicitly since we use the exact sampling distribution
of $Z_T$), but JL lemma also generalizes to other ensembles satisfying certain 
requirements on the decay of the tails \cite{Ward, dasgupta}. 

\begin{figure}[t]
\begin{center}
\epsfig{figure=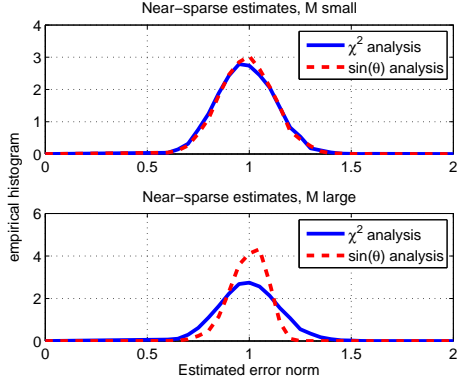,width=2.75in}
\caption{\label{fig:JL_vs_sin_smM} Comparison of $\chi^2$ and $\sin(\theta)$ analysis. 
Given a unit-norm vector $\bdelta$, we obtain $T$ additional measurements, and compute 
our two estimates of $\Vert \bdelta \Vert$. We plot the histogram of the estimates 
over 5000 trials with $N = 250$, $T = 25$, and (a) $M = 0$, (b) $M = 200$.
}
\end{center}
\end{figure}

To compare our analysis in (\ref{eqn:near_sparse_error_v2}), based on $C_T$,  
to the one in (\ref{eqn:JL_analysis}) we note that the latter simply estimates the 
error $\Vert \bdelta \Vert$ as $\Vert \frac{1}{\sqrt{T}} \tilde{A} \bdelta \Vert$, 
where $\tilde{A}$ are the new measurements\footnote{This is essentially the same estimate 
as the one based on JL lemma in \cite{Ward}, as the expected value of $\chi^2_T$ is $T$, 
hence $E[Z_T] = T \Vert \bdelta \Vert_2^2$.}. Now unlike the analysis 
in (\ref{eqn:JL_analysis}), in (\ref{eqn:near_sparse_error_v2}) we require
that the solution at step $M$ is feasible (matches all the measurements) and instead 
we compute the error of projecting $\bdelta$ onto the null-space of $A$ and adjust
it by the expected value of $\frac{1}{\sin(\theta)}$, i.e. we estimate $\Vert \bdelta \Vert$ as 
$\sqrt{ \frac{L}{T}} \Vert A' (A A')^{-1} A \bdelta \Vert$, where $A$ includes all $M+T$ 
measurements. To compare the quality of the two estimates we conducted a simulation with $N = 250$ 
and $T = 25$, and computed the estimates for random unit-norm vectors $\bdelta$. We plot 
the histograms for $M = 0$ and $M = 200$ over $5000$ trials in Figure \ref{fig:JL_vs_sin_smM}. 
In the first case with $M = 0$, we see that both estimates have about the same accuracy (similar 
error distributions), however as $M$ becomes appreciable the approach in 
(\ref{eqn:near_sparse_error_v2}) becomes more accurate. 

\section{Noisy case}
\label{S:noisy}

Next we consider the sequential version of the noisy measurement setting, 
where the observations are corrupted by additive uncorrelated i.i.d. Gaussian
noise with variance $\sigma_n^2$:
\begin{equation}
 y_i = \ba_i' \bx + n_i, ~~~~i \in \{1, .., M\}.
\end{equation}
To solve this problem one can adapt a variety of sparse solvers which 
allow inexact solutions $\hbx^M$ in the sequential setting -- for example matching 
pursuit methods with a fixed number of steps, or the noisy versions of basis 
pursuit. All of these methods have a trade-off between sparsity of the 
desired solution and the accuracy in representing the measurements. In the case
of basis pursuit denoising a regularization parameter $\lambda$ balances these two costs:
\begin{equation}
 \label{eqn:seq_noisy_lp}
 \hbx^M = \arg \min \frac{1}{2} \Vert \by_{1:M} - A^M \bx \Vert_2^2 +
 \lambda_M \Vert \bx \Vert_1.
\end{equation}
For greedy sparse solvers such as matching pursuit and its variants the trade-off
is controlled directly by deciding how many columns of $A$ to use 
to represent $\by$. We are interested in a stopping rule which tells us that $\hbx$ 
is reasonably close to $\bx^*$ for any sparse solver and for any user defined choice of the
trade-off between sparsity and measurement likelihood. We do not discuss the 
question of selecting a choice for the trade-off -- we refer the readers to 
\cite{johnstone_universal, GPSR_wright} and also to \cite{Ward} for a discussion of
how this can be done in a cross-validation setting. 
Now, due to the presence of noise, exact agreement will not occur no matter 
how many samples are taken.  We consider a stopping rule similar to the 
one in Section \ref{S:near_sparse}. In principle, the analysis in
(\ref{eqn:near_sparse_error}) can be extended to the noisy case, but
we instead follow the simplified analysis in Section
\ref{S:near_sparse_simplified}. 

We establish that the reconstruction error can be bounded with high
probability by obtaining a small number of additional samples, and
seeing how far the measurements deviate from $\hat{y}_i = \ba_i'
\hbx^M$. With such a bound one can stop receiving additional
measurements once the change in the solution reaches levels that can
be explained due to noise. The deviations $z_i$ now include
contribution due to noise:
\begin{equation}
 \label{eqn:noisy_error}
 z_i = \hat{y}_i - y_i = \ba_i'( \hbx^M - \bx^*) - n_i.
\end{equation}
Let $Z_T = \sum z_i^2$. Consider the Gaussian measurement ensemble. 
Then $z_i = \ba_i' \bdelta + n_i$, and $\tilde{Z}_T
\triangleq \frac{Z_T}{\Vert \delta \Vert^2 + \sigma_n^2} \sim
\chi^2_T$.  The distribution of $z_i$ is Gaussian with mean zero and
variance $\Vert \bdelta \Vert_2^2 + \sigma_n^2$. Now following a
similar analysis as in previous section we can obtain an estimate of
$\Vert \bdelta \Vert_2^2 + \sigma_n^2$ from a sample of $Z_T$, and
subtracting $\sigma_n^2$ we get an estimate of $\Vert \bdelta \Vert_2^2$.

\begin{figure}[t]
\begin{center}
\epsfig{figure=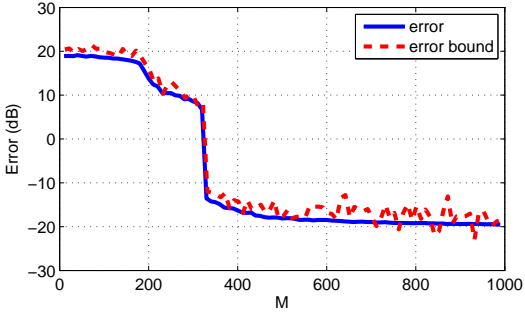,width=3in}
\caption{\label{fig:est_error_noisy} Error estimate in the noisy
case: true error and a $90$-percent confidence bound (dB scale):
$N=1000$, $T=10$, $K = 100$.}
\end{center}
\end{figure}

We show an example in Figure \ref{fig:est_error_noisy} where the
true error appears along with a $90$-percent confidence bound. We have
$N = 1000$, $K = 100$, $T = 10$ and $\sigma_n = 0.01$.  We use
basis pursuit denoising (\ref{eqn:noisy_error}) as our choice for sparse solver, 
and we set $\lambda_M \propto \sqrt{M \log(N)}$ motivated by the universal rule for wavelet 
denoising \cite{johnstone_universal} to account for noise added with additional 
measurements. The bound clearly shows where the sparse 
signal has been recovered up to the noise floor (the signal is sparse with 
$K = 100$ non-zero elements).

\section{Efficient sequential solution}
\label{S:lin_prog}

The main motivation for the sequential approach is to reduce the
number of measurements to as few as possible. Yet, we would also like
to keep the computational complexity of the sequential approach
low. We focus on the $\ell_1$-based formulations here, and show that 
there is some potential of using "memory" in the sequential setting 
for reducing the computational complexity. For the static setting there exists
a great variety of approaches to solve both noiseless and noisy basis 
pursuit (i.e. basis pursuit denoising) in various forms, e.g. \cite{GPSR_wright, 
Boyd_IP_l1, iterative_thresh}. However, instead of re-solving the linear 
program (\ref{eqn:seq_lp}) after each new sample, we would like to use the solution to the
previous problem to guide the current problem. It is known that
interior point methods are not well-suited to take advantage of such
``warm-starts'' \cite{GPSR_wright}. Some methods are able to use
warm-starts in the context of following the solution path in
(\ref{eqn:seq_noisy_lp}) as a function of $\lambda$ \cite{GPSR_wright,
Osborne:homotopy_cont, ICASSP:homotopy}. In that context the solution
path $\hbx(\lambda)$ is continuous (nearby values of $\lambda$ give
nearby solutions) enabling warm-starts. However, once a new
measurement $\ba_i$ is received, this in general makes the previous
solution infeasible, and can dramatically change the optimal solution,
making warm-starts more challenging\footnote{In related work,
\cite{Tropp:row_action_short} proposed to use Row-action methods for
compressed sensing, which rely on a quadratic programming formulation
equivalent to (\ref{eqn:seq_lp}) and can take advantage of sequential
measurements.}.

We now investigate a linear programming approach for warm-starts using
the simplex method to accomplish this in the noiseless case (a similar
strategy can be used with the Dantzig decoder 
\cite{Candes:compressive_sampling} for the noisy case). We can not use the 
solution $\hbx^M$ directly as a starting point for the new
problem at step $M+1$, because in general it will not be feasible. In
the Gaussian measurement case, unless $\hbx^M = \bx^*$, the new
constraint $\ba_{M+1}' \hbx^M = y_{M+1}$ will be violated. One way to
handle this is through a dual formulation\footnote{If at step $M$ the 
optimal dual solution is $\mathbf{p}$, then a feasible solution 
at step $M+1$ is $[\mathbf{p}; 0]$. However, it may not be a basic feasible 
solution.}, but we instead use an augmented primal 
formulation~\cite{Bertsimas:LP}.

First, to model (\ref{eqn:seq_lp}) as a linear program we use the
standard trick: define $x_i^+ = \max(x_i, 0)$, $x_i^- = \max(-x_i,
0)$, and $\bx = \bx^+ - \bx^-$. This gives a linear program in
standard form:
\begin{gather}
  \label{eqn:lp_v1}
  \min \boldsymbol{1}' \bx^+ + \boldsymbol{1}' \bx^- \\ \by_{1:M} =
\left[ A^M ~~-A^M \right] \left[ \begin{smallmatrix}\bx^+ \\ \nonumber
\bx^-
\end{smallmatrix} \right], ~~\mbox{ and}~~ \bx^+, \bx^- \ge 0
\end{gather}

Next we need to add an extra constraint $y_{M+1} = \ba_{M+1}' \bx^+ -
\ba_{M+1}' \bx^-$. Suppose that $\ba_{M+1}' \hbx^M > y_{M+1}$. We add
an extra slack variable $z$ to the linear program, and a high positive
cost $Q$ on $z$. This gives the following linear program:
\begin{gather}
  \label{eqn:lp_v2}
  \min \boldsymbol{1}' \bx^+ + \boldsymbol{1}' \bx^- + Q z \\
\by_{1:M} = \left[ A^M ~~-A^M \right] \left[ \begin{smallmatrix}\bx^+ \\
\nonumber \bx^-
\end{smallmatrix} \right], ~~\mbox{ and}~~ \bx^+, \bx^- \ge 0\\ \nonumber
y_{M+1} = \ba_{M+1}' \bx^+ - \ba_{M+1}' \bx^- - z, ~~\mbox{ and}~~ z
\ge 0
\end{gather}

Now using $\hbx^M$ and $z = \ba_{M+1}' (\hbx^M)^+ - \ba_{M+1}'
(\hbx^M)^- - y_{M+1}$ yields a basic feasible solution to this
augmented problem.  By selecting $Q$ large enough,\footnote{E.g. the 
big-$M$ approach \cite{Bertsimas:LP} suggests treating $Q$ as an 
undetermined value, and assumes that $Q$ dominates when compared to any 
other value.} $z$ will be removed from the optimal basis (i.e. $z$ is set to $0$), 
and the solutions to this problem and the $(M+1)$-th sequential problem 
are the same.

\begin{figure}
  \begin{center}
    \centerline{\epsfig{figure=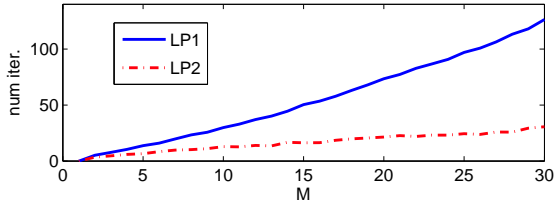,width=3.25in}}
    \caption{\label{fig:lp_init} A comparison of the number of simplex
     iterations when solving (\ref{eqn:seq_lp}) from scratch (LP1) and
     using the solution at step $M-1$ (LP2). We plot the average number of
     iterations vs. $M$, over $100$ trials. }
  \end{center}
  \vspace{-0.5cm}
\end{figure}

We test the approach on an example with $N = 200$, $K=10$, and $100$
trials. In Figure \ref{fig:lp_init} we plot the number of iterations
of the simplex method required to solve the problem (\ref{eqn:seq_lp})
at step $M$ from scratch (LP1) and using the formulation in
(\ref{eqn:lp_v2}) (LP2). To solve (\ref{eqn:lp_v1}) we first have to
find a basic feasible solution, BFS, (phase 1) and then move from it to the
optimal BFS. An important advantage of (\ref{eqn:lp_v2}) is that we
start right away with a BFS, so phase 1 is not required. The
figure illustrates that for large $M$ the approach LP2 is
significantly faster.

We note that recently a very appealing approach for sequential solution in 
the noisy setting has been proposed based on the homotopy continuation idea
\cite{Ghaoui_seq_CS, Romberg_seq_CS}, where a homotopy (a continuous transition) 
is constructed from the problem at step $M$ to the problem at step $M+1$ and 
the piecewise-smooth path is followed. The efficiency of the approach depends 
on the number of break-points in this piecewise-smooth path, but the simulations 
results in the papers are very promising. We also note that \cite{Ghaoui_seq_CS}
proposes an approach to select the trade-off in the noisy case, using 
cross-validation ideas. 

\section{Conclusion and discussion}

This paper presents a formulation for compressed sensing in which the
decoder receives samples sequentially, and can perform computations in
between samples. We showed how the decoder can estimate the error in
the current reconstruction; this enables stopping once the error is
within a required tolerance. Our results hold for any decoding
algorithm, since they only depend on the distribution of the
measurement vectors. This enables ``run-time'' performance guarantees
in situations where a-priori guarantees may not be available, e.g. 
if the sparsity level of the signal is not known, or for
recovery methods for which such guarantees have not been established.

We have studied a number of scenarios including noiseless, noisy,
sparse and near sparse, and involving Gaussian and Bernoulli
measurements, and demonstrated that the sequential approach is
practical, flexible and has wide applicability. A very interesting
problem is to both extend the results to other measurement ensembles,
e.g. for sparse ensembles, and moreover, to go beyond results for
particular ensembles and develop a general theory of sequential
compressed sensing. Furthermore, in many important applications the
sparse signal of interest may also be evolving with time during the
measurement process. Sequential CS with a notion of 'time of a
measurement' is a natural candidate setting in which to explore this
important extension to the CS literature.

We also remark that there is a closely related problem of recovering low-rank 
matrices from a small number of random measurements \cite{Recht_nuclear_norm, 
Montanari_lowRank}, where instead of searching for sparse signals one looks 
for matrices with low-rank. This problem admits a convex 'nuclear-norm' 
relaxation (much akin to $\ell_1$ relaxation of sparsity). Some of our results 
can be directly extended to this setting -- for example if in the Gaussian 
measurement case with no noise there is one-step agreement, then the recovered 
low-rank matrix is the true low-rank solution with probability one. 

Finally we comment on an important question \cite{Weiss_CS,
Seeger_bio} of whether it is possible to do better than simply using
random measurements -- using e.g. experiment design or active learning
techniques. In \cite{Seeger_bio} the authors propose to find a
multivariate Gaussian approximation to the posterior $p(\bx ~| \by)$
where $p(\by ~|~ \bx) \propto \exp( \frac{1}{\sigma^2} \Vert \by - A
\bx \Vert^2)$, and $p(\bx) \propto \exp( -\lambda \Vert \bx \Vert_1)$.
Note that MAP estimation in this model $\hbx = \arg \max_\bx p(\bx ~|~
\by)$ is equivalent to the formulation in (\ref{eqn:seq_noisy_lp}),
but does not provide uncertainties. Using the Bayesian formalism it is
possible to do experiment design, i.e. to select the next measurement
to maximally reduce the expected uncertainty. This is a very exciting
development, and although much more complex than the sequential
approach presented here, may reduce the number of required samples
even further.

\appendices
\section{Derivation of the distribution for $\frac{1}{\sin{\theta}}$}
\label{App:distrib_angles}

Consider $E[ \sin^2(\theta) ] = E[\left(\sum_{i=1}^T h_i^2\right)/\Vert
\bh \Vert_2^2]$. Since $\sum_i E[\frac{h_i^2}{\Vert \bh \Vert_2^2}] =
1$, and each $h_i$ is i.i.d., we have $E[\frac{h_i^2}{\Vert \bh
\Vert_2^2}] = \frac{1}{L}$. In fact $E[\frac{h_i^2}{\Vert \bh
\Vert_2^2}]$ follows a Dirichlet distribution.  Therefore, $E[
\sin^2(\theta) ] = \frac{T}{L}$. 

Using Jensen's inequality with the convex function $\sqrt{1/x}$, $x>
0$, we have $E[ 1/ \sin(\theta) ] \ge \sqrt{\frac{L}{T}}$. 

Now, $E[\frac{1}{\sin^2(\theta)}] = \frac{L-2}{T-2}$ (for $T >
2$). This is true because $
E[\frac{1}{\sin^2(\theta)}] = E \left(\sum_{i=1}^L h_i^2\right)/\left(\sum_{i=1}^T h_i^2\right) = $
$1 + E \left(\sum_{i={T+1}}^L h_i^2 / \sum_{i=1}^T h_i^2\right) = 1 + (L-T) \frac{1}{T-2}$. 
The second term is a product of a $\chi^2$ random variable with $(L-T)$ degrees of 
freedom and an independent inverse-$\chi^2$ distribution with $T$ degrees of freedom: 
$E[\sum_{i={T+1}}^L h_i^2] = L-T$, and 
$E[\frac{1}{\left(\sum_{i=1}^T h_i^2\right)}] = \frac{1}{T - 2}$, see \cite{Kotz:Stat_dist}.
Now $1 + (L-T)/(T-2) = (L-2)/(T-2)$.\\

Finally, using Jensen's inequality with the concave function $\sqrt{x}$, 
$E[\frac{1}{\sin(\theta)} ] \le \sqrt{\frac{T - 2}{L - 2 }}$. 


\bibliographystyle{IEEEtran}
\bibliography{cs_journal}

\end{document}